\documentclass[preprint,12pt]{elsarticle}

\usepackage{indentfirst} 
\usepackage{bm} 
\usepackage{graphicx} 
\usepackage{amsmath} 
\usepackage{bbm}
\usepackage{amsfonts,amssymb}

\newcommand{\be}{\begin{equation}}
\newcommand{\ee}{\end{equation}}
\newcommand \ra {\rightarrow}

\begin{document} 

\begin{frontmatter}



\title{How to Introduce Temperature to the 1 Dimensional Sznajd Model?}


\author{Grzegorz Kondrat}

\address{Institute of Theoretical Physics, University of Wroc{\l}aw, 
    pl. Maxa Borna 9, 50-204 Wroc{\l}aw, Poland}

\ead{gkon@ift.uni.wroc.pl}

\begin{abstract}
We investigate the possibility of introducing the temperature to the one dimensional Sznajd model and propose a natural extension of the original model by including other types of interactions.
We characterise different kinds of equilibria into which the extended system can evolve.  
We determine the consequences of fulfilling the detailed balance condition and we prove that in some cases it is equivalent to microscopic reversibility. We propose a simple definition of the temperature-like quantity that measures the size of fluctuations in the system at equilibrium. The complete list of zero-temperature degenerated cases is provided.  
\end{abstract}

\begin{keyword}
equilibrium \sep temperature \sep detailed balance \sep opinion dynamics

\PACS 64.60.De \sep 05.50.+q \sep 64.70.qd

\end{keyword}

\end{frontmatter}


\newpage
\section{Introduction}
The Sznajd model was originally introduced in \cite{sznajd} in order to describe the mechanisms
of opinion change in the society. The basic element of this approach is the ``social validation",
a social phenomenon relying on the fact that usually two people sharing common opinion 
have much bigger influence on other people in a group than separate individuals.
The model itself and its modifications have found numerous applications in sociophysics 
(see \cite{cas_for_lor_07} for a recent review), marketing 
\cite{schulze,weron2}, finance \cite{weron3} and politics \cite{stauffer,bernardes,sznajd2}.
Apart from interest in social science, finance and politics, it also brings some new ideas 
to physics, as noted by Slanina and Lavicka \cite{slanina}. Some theoretical aspects of the model 
were investigated in \cite{slanina,stauffer2,roehner}. 

The Sznajd model provides a new scheme of interacting between the particles. 
The basic component of the Sznajd model is a lattice of Ising spins.
Each spin is in one of two states: ``up" or ``down", like in the standard Ising model.
Here the orientation of a given spin designates the opinion of an individual described by this spin, 
e.g. voting for or against in some political context.
In the Sznajd model (and in other outflow models in general) 
we have the following dynamics. Within the common framework of random updating some chosen spins 
(their number depends on the variation of the model) influence their outer neighbors, as opposed to
e.g. Glauber dynamics, where two spins affect the spin between them. In the first case we speak about 
{\em outflow dynamics}, as information flows outwards, and in the second case, we have 
{\em inflow dynamics}. Both kinds of models provide microscopic description of macroscopic changes 
in the system, e.g. phase transitions \cite{przejscia_fazowe}. Some dynamical aspects of 
one- and two-dimensional versions of the Sznajd model were considered 
in \cite{weron4,weron5,sznajd-krupa}. Interesting scaling behavior of relaxation times 
for two-dimensional Sznajd model were reported in \cite{kondrat1}. A simple 
underlying dynamical framework that drives global dynamics of the system was proposed for
both Sznajd and Glauber zero-temperature models in \cite{kondrat2}.

In almost all studies of outflow-type models the rules that govern dynamics are deterministic,
and therefore these models are considered as zero-temperature. In the social context 
this means that all people behave in the same, predictable way, which is extremely rare
in real situations.
In general, different types of social response are possible \cite{NDL2000}, such as conformity, 
anticonformity, independence and congruence. These terms are related to various ways of changing  mind in different conditions (e.g. conformity is the core of ,,social validation", as a group tries to make an individual following the group's opinion). 
Certainly human behavior is not predictable 
in 100\% and with some probability one can expect any kind of social response.
Therefore it would be reasonable and interesting to seek a model that incorporates 
some diversity or randomness in human activity. 
On the physical grounds this relates to a question:
{\em How to incorporate noise, or temperature, to the original model in order to allow for 
randomness inherent in many real systems?} Such a development of the Sznajd model will allow us to formulate some general remarks about types of equilibrium that can be reached by the outflow models.

In the original Sznajd model there is no room for the temperature, and so an extension of the model
is needed. Here (in the section two) we propose such extended model. 
Our framework comes from taking into account all possible configurations of small vicinity of each spin (or individual). It is so general, that
it contains some previously examined models \cite{weron4,sanchez,krupa} as special cases. 
In order to introduce the notion of temperature in the model one should examine the equilibrium of the system first (the standard temperature is well defined in equilibrium only). 
The discussion of that issue is provided in the section three.
In the next section we propose a natural candidate for the temperature in the extended model.
Conclusions are drawn in the last (fifth) section.
In the first appendix we present a complete list of special (degenerated) cases of the extended model, for which the dynamics always ceases after some finite time. 
In the second appendix we show the detailed arguments leading from the detailed balance conditions to the constraints (\ref{cond1&2}) and (\ref{cond3&4}).
 
It should be also noticed that the results presented in this paper are rigorous and belong 
to very few contributions that analyze the outflow dynamics analytically \cite{slanina}.

\section{The extended model}
In the original Sznajd model \cite{sznajd} of outflow dynamics 
one considers a system of $L$ Ising spins $S_i=\pm 1$ 
on one dimensional lattice with periodic boundary conditions. In each update 
two nearest neighbors are selected at random. Let us assume that spins $i$ and $i+1$ were selected. 
If they are parallel ($S_i=S_{i+1}$), 
their outer neighbors' spins follow the spins of the selected pair (that is $S_{i-1}=S_{i+2}=S_i$). 
In the case of antiparallel alignment ($S_i\neq S_{i+1}$) two options are considered: 
either the outer spins acquire antiparallel orientation with respect to their closer neighbor
($S_{i-1}=S_{i+1}$ and $S_{i+2}=S_i$) or nothing is changed. 
The latter option was considered to be more natural and was often used in
the literature, but some other rules 
(for antiparallel alignment) were also used, see e.g. \cite{weron4,sanchez,krupa}. 
The dynamics defined in this way is undoubtedly zero-temperature, as no random noise
is present in the system. All initial states approach one of the two final ferromagnetic states 
(all spins up or all spins down) in a finite time. The distribution of the relaxation times 
of this process was investigated in \cite{sznajd-krupa}. 

The original model dynamic rules act only on configurations of the type
$\dots A[BB]A\dots$ (from now on the selected pair of spins, that affect their outer neighbours, we will put in square brackets), where $A=\pm 1$ and $B=\mp 1$ stands for different spin orientations. 
In the cases of antiparallel alignment of the selected spins and in the case of all four spins parallel in a row nothing happens. The natural extension of the model accounts for allowing spin changes in these cases. 
Even though there are some other ways of introducing the noise to the Sznajd model 
(e.g. \cite{sabatelli}), the one proposed here seems the most natural.

We consider a 1D lattice with periodic boundary conditions (as in the original model). 
There are $L$ Ising spins on the lattice, and the dynamics of the model is given by special rules
of changing spins in a single update step. At each step two consecutive spins are chosen at random, and they influence their direct neighbors according to the rules.
Let us consider all possible configurations of 4 consecutive spins (two middle spins in brackets
will control the outcome of the update step). 
We do not consider the external field, so our extended evolution rules must be symmetric 
with the respect to spin reversal ($+ \leftrightarrow -$).  
We consider the action of a selected pair independently in each direction, so only three spins are to be taken into account (two selected and one to be updated).
Thus all different possible elementary cases make up the following list:
($[AA]A$, $[AA]B$, $[AB]A$ and $[AB]B$). 
To determine the dynamics we have to introduce a vector of probabilities ${\bf p}=(p_1,p_2,p_3,p_4)$ 
of flipping the third spin in the configurations below:
\begin{eqnarray}
p_1:[AA]A\ra [AA]B,\label{rules1}\\ 
p_2:[AA]B\ra [AA]A,\\ 
p_3:[AB]A\ra [AB]B,\\ 
p_4:[AB]B\ra [AB]A.\label{rules4}
\end{eqnarray}

The values of these four constants (or parameters) are crucial in the model and govern the dynamics.
For convenience we write the complementary probability: $p_i'=1-p_i$ for $i=1,...,4$, which describes the chance of not flipping the spin in a given configuration.

From the social point of view the parameters $p_i$ denotes the intensity of specific social behavior: non-conformity ($p_1$), conformity ($p_2$) and more subtle cases of changing one's opinion in view of non-conforming nearest and next nearest neighbors ($p_3$ and $p_4$).

\section{Equilibrium}
The notion of equilibrium in statistical physics is fundamental. Many properties of 
the systems are defined only in an equilibrium state (or at least in local 
equilibrium, where the changes of macroscopic quantities are relatively slow). In 
many cases however, systems cannot reach equilibrium due to e.g. existence of 
external forces. But the meaning of equilibrium is sometimes not clear -- there are several non-equivalent  definitions of an equilibrium in the literature. Let us consider the dynamical system with a finite microstate space $\cal{I}$. Let $q_i(t)$ be the probability that the system 
at time $t$ is in the microstate $i\in \cal{I}$. The most basic definition of an 
equilibrium \cite{vankampen} states that the distribution of probability among 
all possible microstates does not vary with time: \be \dot{q}_i(t)=0\mbox{ for all 
} i\in \cal{I}.\label{equilib_0} \ee This condition can be rewritten using the 
notion of transition probabilities between the states. Let $W_{n n'}$ be the 
probability of transition between the microstates: $n'\ra n$ in a single time step 
(or probability density for continuous time). Then the condition (\ref{equilib_0}) 
states that the probability of leaving the state $n$ balances the probability of 
arriving at the state $n$ from all other states: \be \sum_{n'} W_{n n'} 
q_{n'}=\left(\sum_{n'}W_{n' n}\right)q_n. \label{FullBC} \ee

There is also another, more strict approach (see, e.g. \cite{henkel_2008}), that 
defines equilibrium of the system in more rigorous way -- it is required that all 
transitions between each two microstates must balance independently. This 
constraint is known as {\em detailed balance condition} (or DBC) and reads: \be 
W_{n n'} q_{n'}=W_{n' n}q_n\label{dbc} \ee for each pair of the microstates $n$ and 
$n'$. The direct consequence of (\ref{dbc}) is the absence of net probability flow 
along cyclic transitions among any three states $i\ra j\ra k\ra i$ (what, in 
general, can take place in the ``basic" equilibrium  satisfying only the full balance 
conditions (\ref{FullBC})).

There is a fundamental interplay between DBC and reversibility. In the theory of 
Markov processes \cite{kelly_1979} DBC is equivalent to reversibility (but not microscopic reversibility, see below). Let us 
consider a stationary Markov process (all finite-dimensional distributions depend 
on time differences and do not change with time translations). A process is said to 
be {\em reversible} if its finite-dimensional distributions do not change with the time 
reversion (the transformation $t\ra \tau-t$ for some $\tau$). Then the necessary 
and sufficient condition for a stationary Markov process to be reversible is just 
fullfiling the DBC for a collection of positive numbers constituting the 
equilibrium distribution of the process. In statistical mechanics the detailed 
balance appears on a basic level. It is known \cite{thomsen_1953} that the second 
law of thermodynamics (nonnegative entropy production in the isolated system) 
together with the detailed balance is equivalent to the condition of microscopic 
reversibility: 
\be 
W_{n n'}=W_{n' n}
\ee
stating that the transition probabilities between two states do not depend 
on the direction of the transition. In deriving Boltzmann H-theorem for diluted gases DBC appears as an assumption of cross section invariance with respect to time inversion.  It is also known \cite{vankampen} that DBC is always satisfied for all classical physical isolated systems, for which hamiltonian and macroscopic observables are even functions of all momenta.

Even though there is a distinct difference 
between the microscopic reversibility condition and the detailed balance condition 
(the first implies the second, but not the opposite, in general) many authors use them  
interchangeably.

Let us now come back to the extended model and investigate the equilibrium types 
the system can reach. Since we deal with the finite size system, the equilibrium 
(in the ``basic" meaning (\ref{equilib_0})) is always reached in the infinite 
time limit \cite{vankampen}. But there are two general possibilities, depending on the choice 
of the parameters $p_1,\dots,p_4$ of the model. First class is characterized
by the existence of absorbing states -- the dynamics comes to an end after a finite time 
of reaching a microstate that system cannot leave. These cases we call {\em degenerated}, 
as the dynamics stops after a finite time. In this situation the set $\Omega$ of final microstates may consist 
of a single element, or for other parameters' value may contain many final microstates, but
there are no allowed transitions between them.
The full list of degenerated cases with the appropriate values of $p_1,\dots,p_4$ 
is provided in the Appendix A. 

In all other cases that make up the second class the equilibrium state approached by the system has
never vanishing fluctuations and the transition probabilities between the microstates realising given equilibrium state do not vanish.

It is an interesting question what kind of equilibrium is reached by the system depending on the model parameters -- or -- what kind of reversibility is fulfilled by the dynamics of the model.
For the equilibrium from the first class all transitions between microstates realising given equilibrium state vanish, so DBC is satisfied trivially (both sides of the equation (\ref{dbc}) equal zero). For such cases the dynamics is fully reversible at equilibrium in such sense that no change can be present -- the system is frozen forever in both time directions.

On the other hand the issue of satisfying DBC in the second class is by no means evident. One of the aims of this work is to clarify this matter.  The details of the reasoning are provided in the appendix B, here we only sketch the main idea and gotten results. Since DBC forbids the net flow of probability along any cycle of transitions (e.g. $i\ra j\ra k\ra i$ against $i\ra k\ra j\ra i$ for any microstates $i,j,k\in {\cal I}$), the skillful choice of triples $(i,j,k)$ and their analysis leads to some constraints on parameters $p_1,\dots,p_4$. It is argued in the appendix B, that DBC implies the following constraints:
\be
p_1=p_2,\label{cond1&2}
\ee
and
\be
p_3=p_4.\label{cond3&4}
\ee
In order to remain in the second class we must assume further that at least one pair of probabilities does not vanish:
\be
p_1>0 \mbox{ or } p_3>0 \label{non0cond}
\ee
(otherwise we have the degenerated case of number 0 from the appendix A).
In turn assuming (\ref{cond1&2}) and (\ref{cond3&4}) leads directly to DBC, since the obtained conditions (\ref{cond1&2}) and (\ref{cond3&4}) are just the statement of microscopic reversibility.
Thus the extended model is truly time-reversible only for parameters' values satisfying
(\ref{cond1&2})-(\ref{non0cond}) and then time reversibility (or DBC) is equivalent to microscopic reversibility.

\section{The temperature}
After investigating the nature of equilibrium in the extended model we now can safely proceed to discuss the notion of temperature in this context. As a temperature is usually interrelated to the size of fluctuations in an equilibrium state, it would be natural to assign the null temperature to the models from the first class with degenerated dynamics, where no changes are present at the equilibrium state. 

In statistical mechanics the inverse of the temperature is usually defined as an energy derivative of the entropy, but there is no hamiltonian (energy) in the model.
We can define however a temperature-like parameter $T$ in such a simple way that it would measure the mean rate of microstate change at equilibrium. To make it system size independent let us normalize it to one: we define $T$ as the average number of elementary changes of the microstate in a single Monte Carlo step divided by the size of the system. Thus $T=0$ indicates no changes at all and $T=1$ means that the microstate is changing at every possible instant of time.   

In general it is difficult to write the way our temperature $T$ depends on all $p_1,\dots,p_4$ in a closed form. For all cases of the degenerated dynamics we have obviously $T=0$. It appears (from Monte Carlo simulations) that in the cases realising the stronger equilibrium (these from the second class) with $p_1=p_2$ and $p_3=p_4$
the dependence of the temperature on the parameters is particularly simple:
\be
T=\frac{p_1+p_2+p_3+p_4}{4}.
\ee
The highest temperature $T=1$ possible in the model (the highest level of noise) is obtained only for the case $p_1=p_2=p_3=p_4=1$. 
We believe that the proposed way of introducing the temperature to the outflow models will prove useful and enable to investigate such systems from a new perspective.

\section{Conclusions}
In this paper we proposed an extension of the Sznajd model that allows for wider variety of interactions and room for random fluctuations that are observed in real systems. In this extended model we analysed in details the nature of possible equilibria reached by the system and discussed in this context the question of reversibility of the dynamics. It appeared that there are special cases of the model's parameters, for which the dynamics is fully reversible. For the proper description of fluctuations in the model we suggested the temperature-like measure that vanishes for all degenerated cases and acquires the particularly simple form in the case of the reversible dynamics. Applications of such temperature needs further study and we believe that this paper will serve as a good starting point in investigating outflow models from a new perspective.

\appendix

\section{The degenerated cases}
\noindent
Detailed investigation of the rules (\ref{rules1})-(\ref{rules4}) leads one to cases, when the dynamics of the system stops after arriving at some absorbing microstate, from which the system cannot escape. Such cases we call {\em degenerated}, since no further nontrivial dynamics is possible.\\  
 
Here we present the complete list of degenerated cases (the set $\Omega$ consists of all final microstates that cannot be escaped from):\\[-10pt]
\begin{enumerate}
\addtocounter{enumi}{-1}
\item $p_1=0$, $p_2=0$,  $p_3=0$, $p_4=0$\\
    trivial case of no dynamics at all, each configuration is final   \\[-5pt]
\item $p_1>0$, $p_2=0$,  $p_3=0$, $p_4=0$\\   
    $\Omega$ consists of configurations with all clusters of size one and two,\\ 
    e.g. $\{...++-+--+--...\}$\\[-5pt]
\item $p_1=0$, $p_2>0$,  $p_3=0$, $p_4=0$\\
    the case corresponding to the standard Sznajd model, for which $p_2=1$.\\   
    $\begin{array}{rl}
    \Omega=\{&...++++++++...,\\
    &...--------...,\\
    &...-+-+-+-+...\}
    \end{array}
    $ \\[0pt]
\item $p_1=0$, $p_2=0$,  $p_3>0$, $p_4=0$\\   
    $\Omega$ consists of configurations with all clusters of size bigger than one,\\ 
    e.g. $\{...+++++--+++...\}$\\[-5pt]
\item $p_1=0$, $p_2=0$,  $p_3=0$, $p_4>0$\\   
    $\begin{array}{rl}\Omega=\{&...-+-+-+-+...\}\end{array}$\\[-5pt]
\item $p_1>0$, $p_2=0$,  $p_3>0$, $p_4=0$\\   
    $\begin{array}{rl}\Omega=\{&...++--++--...\}\end{array}$ \\[-5pt]
\item $p_1>0$, $p_2=0$,  $p_3=0$, $p_4>0$\\   
    $\begin{array}{rl}\Omega=\{&...-+-+-+-+...\}\end{array}$ \\[-5pt]
\item $p_1=0$, $p_2>0$,  $p_3>0$, $p_4=0$\\   
    $\begin{array}{rl}
    \Omega=\{&...++++++++...,\\
    &...--------...\}
    \end{array}$ \\[5pt]
\item $p_1=0$, $p_2>0$,  $p_3=0$, $p_4>0$\\   
    $\begin{array}{rl}
    \Omega=\{&...++++++++...,\\
    &...--------...,\\
    &...-+-+-+-+...\}
    \end{array}
    $\\[5pt]
\item $p_1=0$, $p_2>0$,  $p_3>0$, $p_4>0$\\
    $\begin{array}{rl}
    \Omega=\{&...++++++++...,\\
    &...--------...\}
    \end{array}$ \\[5pt]
\item $p_1>0$, $p_2>0$,  $p_3=0$, $p_4>0$\\   
    $\begin{array}{rl}\Omega=\{&...-+-+-+-+...\}\end{array}$       
\end{enumerate}

When the lattice size $L$ is an odd number: a) the cases 4, 5, 6 and 10 are no longer degenerated, 
b) in the cases 2 and 8 the antiferromagnetic state does not belong to $\Omega$.
If $L$ is an even number not divisible by 4 the case 5 is no longer degenerated as well.

Let us make a note on the meaning of the above cases on sociological grounds.
The case no. 0 is not interesting -- nothing happens.
The next four cases (1.-4.) describe situation, when only one type of transition is allowed --
one kind of behavior: conformist (the case 2. -- the Sznajd model), pure nonconformist (the case 1.)
or more subtle nonconformist (the cases 3. and 4.). The cases 5.-8. allow for two
different types of behavior that compete with each other, but always one of them finally wins.
The last two cases (9. and 10.) admit three kinds of behavior at a time, but finally 
one of them always wins. Apart from the cases 0., 1., 3. and 5. the system ends in a stable state
of either perfect conformity or total disagreement (each two nearest neighbors do not
agree with each other).

\section{Derivation of constraints (\ref{cond1&2}) and (\ref{cond3&4})}

We now show how satisfying DBC in the extended Sznajd model leads to the constraints (\ref{cond1&2}) and (\ref{cond3&4}). The main idea is to require equality between the rates of cyclic transitions between suitably chosen three microstates in both directions.

Let us consider transitions between the following three microstates:
\begin{eqnarray}
(I): &  ...---(+)++(+)---...\\
(II): & ...---(+)++(-)---...\\ 
(III): &...---(-)++(-)---...
\end{eqnarray}
The dots $\dots$ above describe parts of the configurations that do not contribute to the changes and are of no interest for us.

In the following we calculate transition rates for cyclic changes 
$I\ra II\ra III\ra I$ and $I\ra III\ra II\ra I$ in terms of all probabilities $p_i$.
These transition rates must be equal, since DBC implies that transition rates 
for cyclic changes at equilibrium should not depend on their direction.

Here for clarity we write the spins that are being changed in parentheses.
Let us concentrate for a while on the transition (I) $\ra$ (II).
There are only two ways of choosing two consecutive spins that can affect a given spin: 
they are nearest neighbors either to the left or to the right, the probability of each choice is $1/L$.
In each case in order to obtain the desired change, 
one neighboring spin of the chosen pair must be changed while the other must not. 
For the pair $++$ (written below in brackets) to the left from the spin being changed: 
\begin{equation}
\begin{array}{c}
...---(+)[++](+)---...\\
\downarrow\\
...---(+)[++](-)---...
\end{array}
\end{equation}
the resulting probability is $p_1$ (probability of changing spin to the right of the chosen pair) 
times
$p_1'=1-p_1$ (probability of not changing spin to the left of the chosen pair).
Let us recall that $p_i$'s are defined by the formulae (\ref{rules1})-(\ref{rules4}) 
and $p_i'=1-p_i$.
Similarly, for the choice of the pair $--$ (again in brackets) to the right
from the spin being changed: 
\begin{equation}
\begin{array}{c}
...---(+)++(+)[--]-...\\
\hspace*{83pt}\downarrow\\
...---(+)++(-)[--]-...\\
\end{array}
\end{equation}
the resulting probability is $p_2$ (probability of changing spin to the left of the chosen pair) 
times
$p_1'=1-p_1$ (probability of not changing spin to the right of the chosen pair).
Thus the overall probability of change (I) $\ra$ (II) in one step reads 
\begin{equation}
W_{I\ra II}=(p_1p_1'+p_2p_1')/L.\label{probab1}
\end{equation}
In this way we calculate the following transition probabilities between:
\begin{equation}
\begin{array}{lcl}
W_{II\ra III}&=&(p_2p_1'+p_1p_2')/L,\\ \label{probab2}
W_{III\ra I}&=&p_2^2/L,\\
W_{II\ra I}&=&(p_2p_1'+p_1p_1')/L,\\
W_{III\ra II}&=&(p_1p_1'+p_2p_2')/L,\\
W_{I\ra III}&=&p_1^2/L.
\end{array}
\end{equation}\\

Let us now write DBC for transitions between any pair of the states considered 
and after taking their product we obtain:
\begin{eqnarray}
\label{cycle}
W_{I\ra II}\,q_{II}\,W_{II\ra III}\,q_{III}\,W_{III\ra I}\,q_{I}=\\ 
=W_{I\ra III}\,q_{III}\,W_{III\ra II}\,q_{II}\,W_{II\ra I}\,q_{I},  \nonumber
\end{eqnarray} 
where $q_X$ denotes the probability that the system is at the state $X$ at equilibrium.
Hence, if $q_{I}q_{II}q_{III}\neq 0$,
\begin{eqnarray}
W_{I\ra II}\,W_{II\ra III}\,W_{III\ra I}
=W_{I\ra III}\,W_{III\ra II}\,W_{II\ra I}. \label{cycle1}
\end{eqnarray} 
Inserting probabilities (\ref{probab1}) and (\ref{probab2}) into (\ref{cycle1}) 
and rearranging the terms, we obtain the condition:
\be
p_1'(p_1+p_2)(p_2-p_1)[p_1'(p_2^2+p_1\,p_2+p_1^2)+p_2'\,p_1\,p_2]=0.
\ee

Taking into account that $p_i\geq 0$ and $p_i'\geq 0$, 
the above equation can be satisfied if and only if 
\be
p_1=1 \mbox{ or }p_1=p_2.\label{cond1}
\ee\\

By the same argument, for another choice:
\begin{eqnarray}
(I): &  ...+--(+)++(+)--+...\\
(II): & ...+--(+)++(-)--+...\\ 
(III): &...+--(-)++(-)--+...
\end{eqnarray}
we obtain conditions:
\be
p_2=1 \mbox{ or }p_1=p_2.\label{cond2}
\ee
Taking the conjunction of the conditions (\ref{cond1}) and (\ref{cond2}) we finally
restrict the admissible choice of $p_1$ and $p_2$ to
\be
p_1=p_2.\label{Acond1&2}
\ee\\

In order to obtain conditions for $p_3$ and $p_4$ we consider transitions between the states:
\begin{eqnarray}
(I): &  ...-+-(+)+-(-)+-+...\\
(II): & ...-+-(+)+-(+)+-+...\\ 
(III): &...-+-(-)+-(+)+-+...
\end{eqnarray}
and between the states:
\begin{eqnarray}
(I): &  ...++-(+)+-(-)+--...\\
(II): & ...++-(+)+-(+)+--...\\ 
(III): &...++-(-)+-(+)+--...
\end{eqnarray}
The resulting conditions are
\be
p_3=1 \mbox{ or }p_3=p_4
\ee
for the first set of the states, and
\be
p_4=1 \mbox{ or }p_3=p_4
\ee
for the second set, respectively. This leads to:
\be
p_3=p_4.\label{Acond3&4}
\ee\\


\begin{thebibliography}{33}
\bibitem{sznajd}
K. Sznajd-Weron and J. Sznajd, Int. J. Mod. Phys. C, 11 (2000) 1157
\bibitem{cas_for_lor_07}
C. Castellano, S. Fortunato and V. Loreto, Rev. Mod. Phys., 81 (2009) 591
\bibitem{schulze}
C. Schulze, Int. J. Mod. Phys. C, 14 (2003) 95
\bibitem{weron2}
K. Sznajd-Weron and R. Weron, Physica A, 324 (2003) 437
\bibitem{weron3}
K. Sznajd-Weron and R. Weron , Int. J. Mod. Phys. C, 13 (2002) 115
\bibitem{stauffer}
D. Stauffer, Advances in Complex Systems, 5 (2002) 97
\bibitem{bernardes}
A.T. Bernardes, D. Stauffer and J. Kertesz, Eur. Phys. J. B, 25 (2002) 123
\bibitem{sznajd2}
K. Sznajd-Weron and J. Sznajd, Physica A, 351 (2005) 593
\bibitem{slanina}
F. Slanina and H. Lavicka, Eur. Phys. J. B, 35 (2003) 279
\bibitem{stauffer2}
D. Stauffer and P.M.C. de Oliveira , Eur. Phys. J. B, 30 (2002) 587
\bibitem{roehner}
B.M. Roehner, D. Sornette and J.V. Andersen, Int. J. Mod. Phys. C, 15 (2004) 809
\bibitem{przejscia_fazowe}
M. Mobilia and S. Redner, Phys. Rev. E, 68 (2003) 046106
\bibitem{weron4}
K. Sznajd-Weron, Phys. Rev. E, 66 (2002) 046131
\bibitem{weron5}
K. Sznajd-Weron, Phys. Rev. E, 70 (2004) 037104
\bibitem{sznajd-krupa}
K. Sznajd-Weron and S. Krupa, Phys. Rev. E, 74 (2006) 031109
\bibitem{kondrat1}
G. Kondrat and K. Sznajd-Weron, Phys. Rev. E, 77 (2008) 021127
\bibitem{kondrat2}
G. Kondrat and K. Sznajd-Weron, Phys. Rev. E, 79 (2009) 011119
\bibitem{NDL2000}
P.R. Nail, G. MacDonald and D.A. Levy, Psychological bulletin, 126 (2000) 454
\bibitem{sanchez}
J.R. Sanchez , arXiv:cond-mat/0408518 (2004) 
\bibitem{krupa}
S. Krupa and K. Sznajd-Weron, Int. J. Mod. Phys. C, 16 (2005) 1771
\bibitem{sabatelli}
L. Sabatelli and P. Richmond, Physica A, 334 (2004) 274
\bibitem{vankampen}
N.G. van Kampen, Stochastic processes in physics and chemistry, Amsterdam: North-Holland, 1987
\bibitem{henkel_2008}
M. Henkel, H. Hinrichsen and S. L\"{u}beck, Non-Equilibrium Phase 
Transitions, vol.I: Absorbing Phase Transitions, Springer, 2008 
\bibitem{kelly_1979}
F. Kelly, Reversibility and Stochastic Networks, Chichester: Wiley, 1979
\bibitem{thomsen_1953}
J.S. Thomsen, Phys. Rev., 91 (1953) 1263
\end{thebibliography}
\end{document}